\documentstyle[12pt,aps]{revtex}

\begin{document}

\hsize = 6.5in
\widetext
\draft
\tighten
\topmargin-48pt

\preprint{EFUAZ FT-97-46}

\title{ABOUT  $\gamma^5$ CHIRAL INTERACTIONS OF
MASSIVE PARTICLES IN THE $(1/2,0)\oplus (0,1/2)$
REPRESENTATION\thanks{Presented at the Mexican School on Nuclear
Astrophysics, Guanajuato, Gto., M\'exico, Aug 13-20, 1997.
A shorten version will be published in ``Causality and Locality
in Modern Physics. Proceedings of a Symposium in honour of Jean-Pierre
Vigier", eds. G. Hunter, S. Jeffers and J.-P. Vigier, Kluwer Academic,
Dordrecht, 1998. The complete version is submitted to ``Int. J. Theor.
Phys."}}

\author{{\bf Valeri V. Dvoeglazov}}

\address{
Escuela de F\'{\i}sica, Universidad Aut\'onoma de Zacatecas \\
Apartado Postal C-580, Zacatecas 98068, Zac., M\'exico\\
Internet address:  valeri@cantera.reduaz.mx\\
URL: http://cantera.reduaz.mx/\~~valeri/valeri.htm
}

\date{August 12, 1997}

\maketitle

\begin{abstract}
We argue that the self/anti-self charge conjugate states of
the $(1/2,0)\oplus (0,1/2)$ representation possess the axial
charge. Furthermore, we analyze the recent claims of the $\sim
{\bbox\sigma} \cdot [{\bf A} \times {\bf A}^\ast ]$  interaction terms
for `fermions'. Finally, we briefly discuss the
problem in the $(1,0)\oplus (0,1)$ representation.
\end{abstract}


\section{Introduction}

The Dirac equation and the relevant theory of
charged particles do {\it not} admit the $\gamma^5$ chiral transformation.
The sign in the mass term in the Lagrangian is reversed under this type of
transformations. In the mean time, the chiral transformations play
significant role in our understandings of the nature of weak and strong
interactions, in the problem of (un)existence of monopoles as well.
Many attempts have earlier been done in order to understand the origin of
the chiral (a)symmetry from the first principles, see, e.~g.,~\cite{Lee}.
Recently, the authors of ref.~\cite{Das}  proposed a very
interesting model of the $m$-deformed {\it non-local} chiral
transformations.  But, they indicated at the importance of further study
of chiral transformations and their relevance to the modern physics.
These matters appear to be of use not only from the viewpoint of the
construction of the fundamental theory for neutral particles (which is our
primary purpose), but the consideration of constructs which {\it
admit} the chiral invariance may also be useful for deeper understanding
processes in QCD and other modern gauge models.

In the present article we prove that
massive self/anti-self charge conjugate states in the $(1/2,0)\oplus
(0,1/2)$ representation possess the axial charge (cf. also
the McLennan-Case reformulation~\cite{MLC} of the Majorana
theory~\cite{MAJOR},
refs.~\cite{DVA,ZIINO,DVO} and earlier papers~\cite{DVA0,DVO0}).
Furthermore, we present explicit examples which are relevant to the
viewpoints of M. Markov~\cite{MARKOV},  S. Weinberg~\cite{WEINBERG} and
P. A. M. Dirac~\cite{DIRAC} about the possibility of different equations
for describing particles  of some representation of the Lorentz group
(particularly, of the  $(1/2,0)\oplus (0,1/2)$ representation).  In
the most comprehensive and clear form it is expressed in  the book,
papers and lectures of S.  Weinberg, e.g.,~\cite{W1}: ``The kinematical
classification of particles according to their Lorentz transformation
properties is entirely (for finite mass) determined by their familiar
representation of the rotation group. It has nothing whatever to do with
the choice of one relativistic wave equation rather than another." See
also~\cite{W2,W3}. In our opinion, it is still required to take into
account the issues related to the inversion group, namely, the
theoretical possibility of unconventional representations of
inversions.\footnote{While this type of theories is usually called as
Wigner- (or ``BWW-") type, see~\cite{Wigner,BWW}, the possibility of
unconventional representations of inversions was first indicated by
Soviet mathematical physicists~\cite{Gelfand}, see also
the relevant papers~\cite{Fushchich}.}

\section{Chiral Interactions}

We start from the observation that the Dirac field operator,
which satisfies the Dirac equation\footnote{As opposed to ref.~[3b]
we use the conventional  notation and metric.
Namely, $g^{\mu\nu} = \mbox{diag} (+1, -1, -1, -1)$ and $\gamma$ matrices
are $$\gamma^0 =\pmatrix{0&\openone\cr \openone &0\cr}\, , \,
\gamma^i = \pmatrix{0&-\sigma^i\cr \sigma^i &0\cr}\, , \, \gamma^5 =
\pmatrix{\openone&0\cr 0& -\openone\cr} \, $$
in the Weyl representation. The Pauli charge-conjugation $4\times 4$
matrix is then $$C = e^{i\vartheta_c}
{\cal C} = e^{i\vartheta_c}
\pmatrix{0&\Theta\cr -\Theta
&0\cr}\, , \, \mbox{where}\,\,\, \Theta = \pmatrix{0&-1\cr 1&0\cr}\, .$$
It has the properties
$$C=C^{^T}\, ,\, C^\ast = C^{-1}\, ,$$
$$C^{-1}\gamma^\mu C = -\gamma^{\mu^{\,\ast}}\, ,\,
C^{-1}\gamma^5  C = -\gamma^{5^{\,\ast}}\, .$$ }
\begin{equation}
[ i\gamma^\mu
\partial_\mu -m ] \psi (x^\mu) = 0\, , \label{1} \end{equation} can be
expanded in the following parts:  \begin{mathletters} \begin{eqnarray}
\psi (x^\mu) &=& \psi_\uparrow (x^\mu) + \psi_\downarrow (x^\mu) \,\, ,\\
\psi_\uparrow (x^\mu) &=& \int \frac{d^3 {\bf p}}{(2\pi)^3 2E_p} \left [
u_\uparrow (p^\mu) a_\uparrow (p^\mu) e^{-i\phi} + {\cal C}
u_\uparrow^\ast  (p^\mu) b_\downarrow^\dagger (p^\mu) e^{+i\phi} \right ]
\,\, ,\label{fo11}\\ \psi_\downarrow (x^\mu) &=& \int \frac{d^3 {\bf
p}}{(2\pi)^3 2E_p} \left [ u_\downarrow (p^\mu) a_\downarrow (p^\mu)
e^{-i\phi} - {\cal C} u_\downarrow^\ast  (p^\mu) b_\uparrow^\dagger
(p^\mu) e^{+i\phi} \right ]\, \label{fo12}, \end{eqnarray}
\end{mathletters}
where $\phi = (Et -{\bf x}\cdot {\bf p})/\hbar$.
The charge-conjugate equation is
\begin{equation}
(i\gamma^\mu \partial_\mu -m) C \psi^\dagger (x^\mu) =0\, \label{2};
\end{equation}
and the counterparts of the `field operators' (\ref{fo11},\ref{fo12})
are ($\vartheta_c =0$)
\begin{mathletters}
\begin{eqnarray}
\psi^c (x^\mu) &=& \psi_\uparrow^c (x^\mu) + \psi_\downarrow^c (x^\mu)
\,\, ,\\
\psi_\uparrow^c (x^\mu) &=& \int \frac{d^3 {\bf p}}{(2\pi)^3
2E_p} \left [ u_\uparrow (p^\mu) b_\downarrow (p^\mu)
e^{-i\phi} + {\cal C} u_\uparrow^\ast  (p^\mu)
a_\uparrow^\dagger (p^\mu) e^{+i\phi} \right ] \,\, ,\\
\psi_\downarrow^c (x^\mu) &=& \int \frac{d^3 {\bf p}}{(2\pi)^3 2E_p} \left
[ - u_\downarrow (p^\mu) b_\uparrow (p^\mu) e^{-i\phi} +
{\cal C} u_\downarrow^\ast (p^\mu) a_\downarrow^\dagger (p^\mu)
e^{+i\phi} \right ]\, .
\end{eqnarray} \end{mathletters}
Both $\psi_{\uparrow}$,\, $\psi^c_{\uparrow}$  and $\psi_{\downarrow}$,\,
$\psi^c_{\downarrow}$ can be used to form self/anti-self charge conjugate
{\it field operators} in the coordinate representation after regarding
corresponding superpositions.  For instance,
\begin{mathletters}
\begin{eqnarray}
\Psi^S &=& {\psi_\uparrow + \psi^c_\uparrow \over 2}=
\int \frac{d^3 {\bf p}}{(2\pi)^3 2E_p} \left [ u_\uparrow (p^\mu)
{a_\uparrow + b_\downarrow \over 2} e^{-i\phi} +
{\cal C} u_\uparrow^\ast (p^\mu)
{a_\uparrow^\dagger + b_\downarrow^\dagger \over 2} e^{+i\phi} \right ] \,
,\label{eq01}\\
\Psi^A &=& {\psi_\downarrow - \psi^c_\downarrow \over 2}=
\int \frac{d^3 {\bf p}}{(2\pi)^3 2E_p} \left [ u_\downarrow (p^\mu)
{a_\downarrow + b_\uparrow \over 2} e^{-i\phi} -
{\cal C} u_\downarrow^\ast
(p^\mu) {a_\downarrow^\dagger + b_\uparrow^\dagger \over 2} e^{+i\phi}
\right ] \, ,\label{eq02}\\
\widetilde \Psi^S &=& {\psi_\downarrow +
\psi^c_\downarrow \over 2} = \int \frac{d^3 {\bf p}}{(2\pi)^3 2E_p} \left
[ u_\downarrow (p^\mu) {a_\downarrow - b_\uparrow \over 2} e^{-i\phi} +
{\cal C}
u_\downarrow^\ast (p^\mu) {a_\downarrow^\dagger - b_\uparrow^\dagger
\over 2} e^{+i\phi} \right ] \, ,\label{eq03}\\
\widetilde \Psi^A &=&
{\psi_\uparrow - \psi^c_\uparrow \over 2} = \int \frac{d^3 {\bf
p}}{(2\pi)^3 2E_p} \left [ u_\uparrow (p^\mu) {a_\uparrow - b_\downarrow
\over 2} e^{-i\phi} -
{\cal C} u_\uparrow^\ast (p^\mu) {a_\uparrow^\dagger -
b_\downarrow^\dagger \over 2} e^{+i\phi} \right ] \, .\label{eq04}
\end{eqnarray}
\end{mathletters}
As opposed to K. M. Case~[3b] we introduce the interaction with
the 4-vector potential in the beginning and substitute $\partial_\mu
\rightarrow \nabla_\mu = \partial_\mu -i e A_\mu$ in the equation
(\ref{1}). For the sake of generality we assume that the 4-vector
potential is a {\it complex} field $A_\mu = C_\mu +i B_\mu$, what is the
extension of this concept comparing with the usual quantum-field
consideration.\footnote{In the modern textbooks on the classical/quantum
field theory the 4-vector potential in the coordinate representation is
usually the {\it real} function(al).  We still note that different choices
of a) relations between the left- and right- parts of the momentum-space
bispinors; b) relations between creation and annihilation operators in the
field operator; and  c) metrics would induce ones to change this conclusion
for interactions of various field configurations which one considers.}
Following the logics of refs.~\cite{MLC,ZIINO} (the separation
of different {\it chirality} sub-spaces) we should consider additional
equations for $\gamma^5 \psi_{\uparrow\downarrow}$ and $\gamma^5
\psi_{\uparrow\downarrow}^c$, i.e., the following set
\begin{mathletters}
\begin{eqnarray}
&&\left [i\gamma^\mu (\partial_\mu -i e C_\mu +eB_\mu)
-m \right ] (\psi_\uparrow + \psi_\downarrow) = 0\, ,\label{eq1}\\
&&\left [i\gamma^\mu
(\partial_\mu +i e C_\mu +eB_\mu) -m \right ] (\psi_\uparrow^c +
\psi_\downarrow^c) = 0\, ,\label{eq2}\\
&&\left [i\gamma^\mu (\partial_\mu -i e
C_\mu +eB_\mu) +m \right ] \gamma^5 (\psi_\uparrow + \psi_\downarrow)
= 0\, ,\label{eq3}\\
&&\left [i\gamma^\mu  (\partial_\mu +i e C_\mu +eB_\mu)
+m \right ] \gamma^5 (\psi_\uparrow^c + \psi_\downarrow^c) = 0
\, . \label{eq4}
\end{eqnarray} \end{mathletters}
Due to symmetries of the Dirac equations
one can proceed in various ways. For instance, let us introduce the
following linear combinations
\begin{mathletters} \begin{eqnarray} \psi_1
&=& \psi_\uparrow^c - \gamma^5 \psi_\downarrow\, ,\quad \psi_2 =
\psi_\downarrow + \gamma^5 \psi_\uparrow^c\, ,\\ \psi_3 &=&
\psi_\downarrow^c + \gamma^5 \psi_\uparrow\, ,\quad \psi_4 =
\psi_\uparrow - \gamma^5 \psi_\downarrow^c\, , \end{eqnarray}
\end{mathletters}
which can be used to represent solutions of Eqs.
(\ref{eq1}-\ref{eq4}).  Then we proceed with simple algebraic
transformations of the set (\ref{eq1}-\ref{eq4}) to obtain ($\widetilde
\nabla_\mu \equiv \partial_\mu +eB_\mu $)
\begin{mathletters}
\begin{eqnarray}
i\gamma^\mu \widetilde \nabla_\mu (\psi_1 - \gamma^5
\psi_4) - e\gamma^\mu C_\mu (\gamma^5 \psi_2 +\psi_3) -m (\gamma^5
\psi_2 +\psi_3) &=& 0\, ,\label{eq11}\\
i\gamma^\mu \widetilde \nabla_\mu
(\gamma^5\psi_2 + \psi_3) -e\gamma^\mu C_\mu (\psi_1 -\gamma^5 \psi_4)
-m (\psi_1 -\gamma^5 \psi_4) &=& 0\, .\label{eq12} \end{eqnarray}
\end{mathletters}
Other two equations are obtained after multiplying
(\ref{eq11},\ref{eq12}) by the $\gamma^5$ matrix.
For the first sight
one could conclude that we would obtain different physical excitations
(due to mathematically different dynamical equations with different
interactions) depending on constraints which we impose on functions
$\psi_{1,2,3,4}$.  Let us impose $\widetilde \Psi^S =0$ and $\widetilde
\Psi^A =0$, see Eqs.  (\ref{eq03},\ref{eq04}).  They are considered here
to be equivalent to either the constraints on the creation/annihilation
operators\footnote{Of course, this is true only if one works with IR
representations of the Wigner-Jordan (anti)commutation rules.} $a_\uparrow
(p^\mu) = b_\downarrow (p^\mu)$ and $a_\downarrow (p^\mu) = b_\uparrow
(p^\mu)$ or the constraints $\psi_\uparrow = \psi_\uparrow^c \equiv
\psi^s$ and $\psi_\downarrow = -\psi_\downarrow^c \equiv
\psi^a$.\footnote{As opposed to the above, one can wish to put the
constraints $\Psi^S=\Psi^A=0$ (or even more general ones), which are
considered  to be equivalent to $a_\uparrow =-b_\downarrow$ and
$a_\downarrow =-b_\uparrow$.  Thus, one can reformulate the formulas in
the rest of the paper. In my opinion, the physical content,  which is
relevant to the aims of the {\it present article}, will not be changed.
So, the constraints are used {\it here} only for the purposes of
simplicity and clarity.} The functions $\psi_{1,2,3,4}$ become to be
interrelated by the conditions
\begin{equation} \psi_1 = \psi^s -\gamma^5
\psi^a\, ,\,\,\, \psi_2 = \psi^a +\gamma^5 \psi^s\, , \psi_3 \equiv
\gamma^5 \psi_1\, , \,\,\, \psi_4 \equiv \gamma^5 \psi_2\, .
\end{equation} It is the simple procedure to show that $\psi_1$ presents
itself self-charge conjugate field and $\psi_2$, the anti-self charge
conjugate field.\footnote{The operator of the charge conjugation and the
chirality $\gamma^5$ operator (chosen as above) are the {\it
anti-commuting} operators.} As the result one obtains \begin{mathletters}
\begin{eqnarray} i\gamma^\mu D_\mu^\ast \psi_1 -m \gamma^5 \psi_2 &=& 0\,
,\label{two1}\\ i\gamma^\mu D_\mu \psi_2 +m \gamma^5 \psi_1 &=& 0\,
,\label{two2} \end{eqnarray} \end{mathletters} where the lengthening
derivative is now defined $$D_\mu = \partial_\mu -ie\gamma^5 C_\mu
+eB_\mu\, .$$ Equations for the Dirac conjugated counterparts of
$\psi_{1,2}$ read \begin{mathletters} \begin{eqnarray} i\partial_\mu
\overline \psi_1 \gamma^\mu -m \overline\psi_2 \gamma^5 &=& 0\, ,\\
i\partial_\mu \overline \psi_2 \gamma^\mu +m \overline\psi_1 \gamma^5 &=&
0\, .  \end{eqnarray} \end{mathletters} One can propose the Lagrangian for
free fields $\psi_{1,2}$ and their Dirac conjugates (cf.  with the concept
of the extra Dirac equations in ref.~[9d] and  with the spin-1 case,
ref.~\cite{DVOEG}):\footnote{At this point we still leave the room for
other kinds of the Lagrangians describing self/anti-self charge conjugate
states, see below and cf.~\cite{DVO}.} \begin{eqnarray} L^{free} &=&
{i\over 2} \left [ \overline \psi_1 \gamma^\mu \partial_\mu \psi_1
-\partial_\mu \overline\psi_1 \gamma^\mu \psi_1 +\overline\psi_2
\gamma^\mu \partial_\mu \psi_2 - \partial_\mu \overline \psi_2 \gamma^\mu
\psi_2 \right ] + \nonumber\\ &-& m\left [\overline \psi_1 \gamma^5 \psi_2
-\overline \psi_2 \gamma^5 \psi_1 \right ]\, ; \end{eqnarray} and the
terms of the interaction:  \begin{equation} L^{int} =   - e (\overline
\psi_1 \gamma^\mu \gamma^5 \psi_1 -\overline \psi_2 \gamma^\mu\gamma^5
\psi_2 ) C_\mu + ie (\overline \psi_1 \gamma^\mu \psi_1 +\overline \psi_2
\gamma^\mu \psi_2 ) B_\mu \end{equation} The conclusion that
self/anti-self charge conjugate can possess the axial charge (of opposite
values) is {\it in accordance} with the conclusions of
refs.~\cite{ZIINO,DVO} and with the old ideas of R. E.
Marshak~\cite{MARSHAK}.  It is remarkable feature of this model that we
did {\it not} assume that self/anti-self charge conjugate fields are
massless.

One can come to this conclusion on using another way of speculations.
Equations (\ref{1},\ref{2}) with interaction can be presented in
two-component form ($\psi = \mbox{column} (\phi\quad\chi )$
and $\sigma^\mu = (\openone_{2\times 2}, -\sigma^i)$):
\begin{mathletters} \begin{eqnarray}
&&i\sigma^\mu \nabla_\mu \chi -m\phi  = 0\, ,\\
&&i\widetilde\sigma^\mu \nabla_\mu \phi -m\chi  = 0\,,\\
&&i\sigma^\mu \nabla_\mu^\ast (-\Theta\phi^\dagger)
-m (\Theta\chi^\dagger)  = 0\, ,\\
&&i\widetilde\sigma^\mu \nabla_\mu^\ast (\Theta\chi^\dagger)
-m (-\Theta\phi^\dagger)  = 0\, ,
\end{eqnarray}
\end{mathletters}
with the hermitian conjugation acting on the $q$- numbers
(it acts on the $c$- numbers as the complex conjugation).
Introducing other bispinors
\begin{equation}
{\cal R}^{S,A} = \pmatrix{\phi\cr \mp\Theta \phi^\dagger\cr}\,\, ,\quad
{\cal L}^{S,A} = \pmatrix{\pm\Theta\chi^\dagger\cr \chi\cr}\,\, ,
\end{equation}
and combining the second and the third equation, and then, the first and
the fourth equations, one can arrive at the equations for new bispinors:
\begin{mathletters}
\begin{eqnarray}
i\gamma^\mu D_\mu \, {\cal R}^{S,A} - m \, {\cal L}^{S,A} &=&0\,
,\label{m1}\\
i\gamma^\mu D_\mu^\ast \, {\cal L}^{S,A} - m\, {\cal
R}^{S,A} &=&0\, \label{m2}.
\end{eqnarray} \end{mathletters}
After taking
into account relations between ${\cal R}^S$ and ${\cal R}^A$ (and between
${\cal L}^S$ and ${\cal L}^A$) we can obtain two sets:
\begin{equation}
i\gamma^\mu D_\mu^\ast \,{\cal L}^S -m\gamma^5 \,{\cal R}^A =0\, ,\quad
i\gamma^\mu D_\mu\, {\cal R}^A +m\gamma^5 \, {\cal L}^S=0\, ;\label{110}
\quad
\end{equation}
and/or
\begin{equation} i\gamma^\mu D_\mu \,{\cal R}^S
+m\gamma^5 \, {\cal L}^A =0\, ,\quad i\gamma^\mu D_\mu^\ast \, {\cal L}^A
-m\gamma^5 \, {\cal R}^S=0\, .\quad\label{120}
\end{equation}
They are precisely the
equations which we obtained before (cf. (\ref{two1},\ref{two2}) and the
equations multiplied by the $\gamma^5$ matrix).  If we now impose the
Majorana {\it anzatz}\,\footnote{$\alpha$ is an arbitrary phase factor. It
is easy to note that in the case of $\alpha=0$ the Majorana {\it anzatz}
results in ${\cal R}^S = +{\cal L}^S$ and ${\cal R}^A = - {\cal L}^A$.
But, in the case $\alpha = \pi$ one obtains ${\cal R}^S = -{\cal L}^S$ and
${\cal R}^A = + {\cal L}^A$.} $\phi = e^{i\alpha}\Theta\chi^\dagger$ on
all four equations we obtain that the $\gamma^5$ interaction terms seem to
disappear.  Due to our previous research~\cite{DVA,ZIINO,DVO}, which was
based on other postulates (see also below), we are sure in the necessity
of modifications of the Dirac theory for neutral particles and in the
presence of the $\gamma^5$ interactions. In fact, Majorana {\it anzatz}
(e.~g., with $\alpha =0$) is connected with the interrelations
between field operators $\psi_{1,2}$ above. So, we may loose some
information. How to solve the problem rigorously? See below.

Further arguments in aid of our reasoning are given by  several constructs
which appeared recently~\cite{ZIINO,DVA,DVO0,DVO}. The Ahluwalia
reformulation of the McLennan-Case construct was presented in
1994~\cite{DVA}. The following type-II spinors have been defined in the
momentum space:
\begin{eqnarray} \lambda^{S,A} (p^\mu) =
\pmatrix{\zeta_\lambda\Theta_{[j]} \phi_{_L}^\ast (p^\mu)\cr \phi_{_L}
(p^\mu)\cr}\, \, ,\quad \rho^{S,A} (p^\mu) = \pmatrix{\phi_{_R} (p^\mu)\cr
(\zeta_\rho \Theta_{[j]})^\ast \phi_{_R}^\ast (p^\mu)\cr}\, .
\end{eqnarray}
In our choice of the operator of the
charge conjugation ($\vartheta_c = 0$) the phase factors
$\zeta_{\lambda , \rho}$ are defined as $\pm 1$, for $\lambda^S$
($\rho^S$), and $\mp 1$, for $\lambda^A$ ($\rho^A$), respectively. One can
find relations between the type-II spinors and the Dirac spinors. They
are listed here
\begin{mathletters} \begin{eqnarray}
\lambda^S_\uparrow
(p^\mu) &=&  +\rho^S_\downarrow (p^\mu) =+ {1-\gamma^5 \over 2} u_\uparrow
(p^\mu) +{1+\gamma^5 \over 2} u_\downarrow (p^\mu) \, ,\label{r1}\\
\lambda^S_\downarrow (p^\mu) &=& -\rho^S_\uparrow (p^\mu) =
-{1+\gamma^5 \over 2} u_\uparrow (p^\mu) +{1-\gamma^5 \over 2}
u_\downarrow (p^\mu) \, ,\label{r2}\\
\lambda^A_\uparrow (p^\mu) &=& -\rho^A_\downarrow (p^\mu) =
+ {1-\gamma^5 \over 2} u_\uparrow (p^\mu) -{1+\gamma^5 \over 2}
u_\downarrow (p^\mu) \, ,\label{r3}\\
\lambda^A_\downarrow (p^\mu) &=& +\rho^A_\uparrow (p^\mu) = + {1+\gamma^5
\over 2} u_\uparrow (p^\mu) +{1-\gamma^5 \over 2} u_\downarrow (p^\mu) \,
.\label{r4}
\end{eqnarray} \end{mathletters}
The positive-energy solutions
are assumed in ref.~\cite{DVA} to be presented by, e.~g., self charge
conjugate $\lambda^S$ spinors, the negative-energy solutions, by
anti-self charge conjugate $\lambda^A$ spinors:\footnote{Let me remind
that the sign of the phase in the field operator is considered to be
invariant if we restrict ourselves by the orthochroneous proper Poincar\'e
group.  This fact has been used at the stage of writing the dynamical
equations (\ref{1100},\ref{1200},\ref{1300},\ref{1400}), see
below.}$^{,}$ \footnote{Field operators in this construct may be not
self/anti-self charge conjugate operator. So, the notation ($S,A$)
used in the formulas in the coordinate space indicates {\it only} the
presence of self/anti-self charge conjugate {\it states} and does not
refer to the properties of the field operator.}
\begin{eqnarray} \nu
(x^\mu) &=& \lambda^S (x^\mu) +\lambda^A (x^\mu) \equiv \label{fo1} \\
&\equiv& \int \frac{d^3 {\bf p}}{(2\pi)^3} \, {1\over 2p_0} \sum_\eta
\left [ \lambda^S_\eta (p^\mu) c_\eta (p^\mu) \exp (-ip\cdot x)
+\lambda_\eta^A (p^\mu) d_\eta^\dagger (p^\mu) \exp (+ip\cdot x) \right
]\,\, . \nonumber \end{eqnarray} Of course, one can construct the field
operator composed of $\rho^{S,A}$ bispinors, e.~g., \begin{eqnarray}
\widetilde\nu (x^\mu) &=& \rho^A (x^\mu) +\rho^S (x^\mu) \equiv
\label{fo2}\\
&\equiv& \int \frac{d^3 {\bf p}}{(2\pi)^3} \, {1\over
2p_0} \sum_\eta \left [ \rho^A_\eta (p^\mu) e_\eta (p^\mu) \exp
(-ip\cdot x) +\rho_\eta^S (p^\mu) f_\eta^\dagger (p^\mu) \exp (+ip\cdot
x) \right ]\,\, .\nonumber
\end{eqnarray}
One of surprising features of this construct~\cite{DVA,DVO} is the fact
that dynamical equations take eight-component form from the beginning. As
shown there the equations for  self/anti-self charge conjugate states
read:\footnote{$\vartheta_c =0$ again. The sign in the mass term depends
on this phase factor.}
\begin{mathletters} \begin{eqnarray} i \gamma^\mu
\partial_\mu \lambda^S (x^\mu) + m \rho^A (x^\mu ) &=& 0 \quad,
\label{1100}\\ i \gamma^\mu \partial_\mu \rho^A (x^\mu) + m \lambda^S
(x^\mu) &=& 0 \quad; \label{1200} \end{eqnarray} \end{mathletters}
and
\begin{mathletters} \begin{eqnarray} i \gamma^\mu \partial_\mu \lambda^A
(x^\mu) - m \rho^S (x^\mu) &=& 0\quad,\\ \label{1300} i \gamma^\mu
\partial_\mu \rho^S (x^\mu) - m \lambda^A (x^\mu) &=& 0\quad.
\label{1400} \end{eqnarray} \end{mathletters}
They can be written in the
8-component form as follows (see formulas (21) in~\cite{DVO}
for $\Gamma$ matrices):
\begin{mathletters} \begin{eqnarray} \left
[i \Gamma^\mu \partial_\mu + m\right ] \Psi_{(+)} (x^\mu) &=& 0\quad,
\label{psi1}\\ \left [i \Gamma^\mu \partial_\mu - m\right ] \Psi_{(-)}
(x^\mu) &=& 0\quad, \label{psi2} \end{eqnarray} \end{mathletters} where
\begin{eqnarray}
\Psi_{(+)} (x^\mu) = \pmatrix{\rho^A (x^\mu)\cr \lambda^S
(x^\mu)\cr}\quad,\quad
\Psi_{(-)} (x^\mu) = \pmatrix{\rho^S (x^\mu)\cr
\lambda^A (x^\mu)\cr}\quad.
\end{eqnarray}
One can reveal the possibility of the $\gamma^5$ phase
transformations~\cite{DVO}.
The Lagrangian~\cite[Eq.(24)]{DVO}, which (like in the Dirac
construct) becomes to be equal to zero on the solutions of the dynamical
equations\footnote{The overline implies the Dirac conjugation.}
\begin{eqnarray}
{\cal L} &=& {i\over 2} \left [
\overline{\lambda^S} \gamma^\mu \partial_\mu \lambda^S -
(\partial_\mu \overline{\lambda^S} ) \gamma^\mu \lambda^S +
\overline{\rho^A} \gamma^\mu \partial_\mu \rho^A -
(\partial_\mu \overline{\rho^A}) \gamma^\mu \rho^A +\right.\nonumber\\
&+&\left.\overline{\lambda^A} \gamma^\mu \partial_\mu \lambda^A -
(\partial_\mu \overline{\lambda^A} ) \gamma^\mu \lambda^A +
\overline{\rho^S} \gamma^\mu \partial_\mu \rho^S -
(\partial_\mu \overline{\rho^S}) \gamma^\mu \rho^S \right ] +
\nonumber\\
&+& m \left [ \overline{\lambda^S} \rho^A + \overline{\rho^A} \lambda^S
-\overline{\lambda^A} \rho^S -\overline{\rho^S} \lambda^A \right ]\,
\end{eqnarray}
is invariant with respect to the phase transformations:
\begin{mathletters}
\begin{eqnarray}
\lambda^\prime (x^\mu)
\rightarrow (\cos \alpha -i\gamma^5 \sin\alpha) \lambda
(x^\mu)\quad,\label{g10}\\
\overline \lambda^{\,\prime} (x^\mu) \rightarrow
\overline \lambda (x^\mu) (\cos \alpha - i\gamma^5
\sin\alpha)\quad,\label{g20}\\
\rho^\prime (x^\mu) \rightarrow  (\cos \alpha +
i\gamma^5 \sin\alpha) \rho (x^\mu) \quad,\label{g30}\\
\overline \rho^{\,\prime} (x^\mu) \rightarrow  \overline \rho (x^\mu)
(\cos \alpha + i\gamma^5 \sin\alpha)\quad.\label{g40}
\end{eqnarray}
\end{mathletters}
Obviously, the 4-spinors $\lambda^{S,A} (p^\mu)$ and $\rho^{S,A}
(p^\mu)$ remain in the space of self/anti-self charge conjugate
states.\footnote{Usual phase transformations like that applied to
the Dirac field will destroy self/anti-self charge conjugacy.
The origin lies in the fact that the charge conjugation operator
is {\it not} a linear operator and it includes the operation of complex
conjugation.} In terms of the field functions $\Psi_{(\pm)} (x^\mu)$ the
transformation formulas recast as follows ($\L^5 = \mbox{diag}
(\gamma^5\quad -\gamma^5)$ and $\Psi_{(\pm)} \equiv
\Psi^\dagger_{(\pm)} \Gamma^0$)
\begin{mathletters} \begin{eqnarray}
\Psi^{\,\prime}_{(\pm)} (x^\mu) \rightarrow \left ( \cos \alpha + i \L^5
\sin\alpha \right ) \Psi_{(\pm)} (x^\mu)\quad,\label{g1}\\
\overline\Psi_{(\pm)}^{\,\prime} (x^\mu) \rightarrow \overline \Psi_{(\pm)}
(x^\mu) \left ( \cos \alpha - i \L^5 \sin\alpha \right )\quad.\label{g2}
\end{eqnarray} \end{mathletters}
Let us proceed further  with  the local gradient
transformations (gauge transformations) in the Majorana-Ahluwalia
construct. When we are interested in them one must introduce the
compensating field of the 4-vector potential~\cite{DVO}
\begin{mathletters} \begin{eqnarray}
&& \partial_\mu
\rightarrow \nabla_\mu = \partial_\mu + ie \L^5 A_\mu\quad,\\ &&
A_\mu^\prime (x) \rightarrow A_\mu (x) - {1\over e} \,\partial_\mu \alpha
\quad.  \end{eqnarray} \end{mathletters}
Therefore, equations describing interactions of the $\lambda^{S}$ and
$\rho^{A}$  with 4-vector potential are the following
\begin{mathletters} \begin{eqnarray}
i\gamma^\mu \partial_\mu \lambda^S (x^\mu)
+e\gamma^\mu \gamma^5 A_\mu \lambda^S (x^\mu)+ m\rho^A (x^\mu) &=&
0\quad,\label{i1}\\
i\gamma^\mu \partial_\mu \rho^A (x^\mu) -e\gamma^\mu
\gamma^5 A_\mu \rho^A (x^\mu) + m\lambda^S (x^\mu) &=& 0\quad.\label{i2}
\end{eqnarray} \end{mathletters}
The second-order equations follow
immediately form the set (\ref{i1},\ref{i2})\footnote{The case of
$\lambda^A$ and $\rho^S$ is similar.}
\begin{mathletters}
\begin{eqnarray}
\left \{\left (i \widehat \partial -e \widehat A
\gamma^5 \right ) \left (i \widehat \partial +e \widehat A \gamma^5
\right ) -m^2 \right \} \lambda^S (x^\mu) &=& 0\quad,\\
\left \{\left (i
\widehat \partial +e \widehat A \gamma^5 \right ) \left (i \widehat
\partial -e \widehat A \gamma^5 \right ) -m^2 \right \} \rho^A (x^\mu) &=&
0\quad;
\end{eqnarray} \end{mathletters}
with the notation being used: $\widehat A \equiv
\gamma^\mu A_\mu = \gamma^0 A^0 - ({\bbox \gamma} \cdot {\bf A})$.
After algebraic transformations in a spirit of~\cite{Ryder,Itzyk}
one obtains
\begin{mathletters} \begin{eqnarray}
\left \{ \Pi_\mu^+ \Pi^{\mu\,+}
-m^2 + {e\over 2} \gamma^5 \Sigma^{\mu\nu} F_{\mu\nu} \right \}
\lambda^{S,A} (x^\mu) &=& 0\quad,\label{ii1}\\
\left \{ \Pi_\mu^-
\Pi^{\mu\,-} -m^2 - {e\over 2} \gamma^5 \Sigma^{\mu\nu} F_{\mu\nu} \right
\} \rho^{A,S} (x^\mu) &=& 0\quad,\label{ii2}
\end{eqnarray}
\end{mathletters}
where  the `covariant derivative' operators acting in
the $(1/2,0)\oplus (0,1/2)$ representation are now defined
\begin{equation}
\Pi_\mu^\pm = {1\over i} \partial_\mu \mp e\gamma^5
A_\mu\quad,
\end{equation}
and
\begin{equation}
\Sigma^{\mu\nu} = {i\over
2} \left [ \gamma^\mu\, ,\, \gamma^\nu \right ]_- \quad.
\end{equation}
Thus, we see that the second-order equations for the particles
described by the field operator $\nu (x^\mu)$ (Eq.  (46) in~\cite{DVA}
and Eq. (\ref{fo1}) of this paper), which interact with the 4-vector
potential, have the same form for positive- and negative-energy parts.
The same is true in the case of the use of the field operator composed
from $\rho^A$ and $\rho^S$. One can see the difference with the Dirac
case; namely, the presence of $\gamma^5$ matrix in the ``Pauli term" and
in the lengthening derivatives. Next, we are able to decouple the set
(\ref{ii1},\ref{ii2}) for the up- and down- components
of the bispinors in the coordinate representation.  For instance, the up-
and the down- parts of the $\nu^{^{DL}} (x) =\mbox{column} (\xi \quad
\eta )$ interact with the vector potential in the following manner:
\begin{eqnarray} \cases{\left
[\pi_\mu^+ \pi^{\mu\,+} -m^2 +{e\over 2} \sigma^{\mu\nu}
F_{\mu\nu} \right ] \xi (x^\mu)=0\quad, &\cr
\left [\pi_\mu^- \pi^{\mu\,-} -m^2
-{e\over 2} \widetilde\sigma^{\mu\nu} F_{\mu\nu} \right ] \eta (x^\mu)
=0\quad, &\cr}\label{iii}
\end{eqnarray}
where already one has $\pi_\mu^\pm =
i\partial_\mu \pm eA_\mu$, \, $\sigma^{0i} = -\widetilde\sigma^{0i} =
i\sigma^i$, $\sigma^{ij} = \widetilde\sigma^{ij} = \epsilon_{ijk}
\sigma^k$.  Of course, introducing the operator composed of the $\rho$
states one can write corresponding equations for its up- and down-
components  and, hence, restore the Feynman-Gell-Mann
equation~\cite[Eq.(3)]{Feynman} and its charge conjugate
(if one considers that $A^\mu$ and $F^{\mu\nu}$ are the real
fields).\footnote{One can connect the Feynman-Gell-Mann field with $\nu
(x^\mu)$ and $\widetilde \nu (x^\mu)$ defined in (\ref{fo1},\ref{fo2}).
For instance,
\begin{mathletters} \begin{eqnarray} \Psi^{FGM} &=&
{1+\gamma^5 \over 2} \widetilde\nu \pm {1-\gamma^5 \over 2} \nu\,\, ,\\
(\Psi^{FGM})^c &=& {1+\gamma^5 \over 2} \nu \pm {1-\gamma^5 \over 2}
\widetilde\nu \,\, .  \end{eqnarray} \end{mathletters} But the signs are
not fixed in the framework of this consideration (due to the fact that the
Feynman-Gell-Mann equations are of the second order and the left-hand side
operator (see Eq. (3) in~\cite{Feynman}) commutes with the $\gamma^5$
matrix.} In fact, this way would lead us to the consideration which is
identical to the recent papers~\cite{Robson}.  It was based on the
linearization procedure for 2-spinors, which is similar to that used by
Feshbach and Villars~\cite{Fesh} in order to deduce the Hamiltonian form
of the Klein-Gordon equation.  Some insights in the interaction issues
with the 4-vector potential in the eight-component equation have been made
there:  for instance, while explicit form of the wave functions slightly
differ from the Dirac case, the hydrogen atom spectrum is the same to that
in the usual Dirac theory~\cite[p.66,74-75]{Itzyk}.  Next, like in the
paper~\cite{Giesen} the equations of~\cite{Robson} presume a
non-CP-violating\footnote{This is possible due to the Wigner ``doubling"
of the components of the wave function.} electric dipole moment of the
corresponding states.

Next, by  using the relations (\ref{r1}-\ref{r4}) one can deduce how is
the $\nu$ operator, which was given by D. V.  Ahluwalia, connected with
the Dirac field operator and its charge conjugate. In the particular case
when $(a_\downarrow + b_\uparrow)/2 = (a_\uparrow -
b_\downarrow)/2 \equiv c_\uparrow = d_\downarrow$ and $(a_\uparrow
+b_\downarrow)/2 = (a_\downarrow - b_\uparrow)/2 \equiv c_\downarrow =
d_\uparrow$ one has
\begin{equation}
\nu (x^\mu) = +{1\over 2}
(\psi_\downarrow (x^\mu) - \psi_\uparrow^c (x^\mu)) - {\gamma^5 \over 2}
(\psi_\uparrow (x^\mu) +\psi_\downarrow^c (x^\mu) )\, .\label{nu1}
\end{equation}
The operator composed of $\rho$ spinors
is then expressed\footnote{Of course, certain relations between
creation/annihilation operators of various field operators are
again assumed.}
\begin{equation}
\widetilde \nu (x^\mu) = +{1\over 2} (\psi_\downarrow (x^\mu) +
\psi_\uparrow^c (x^\mu))
+{\gamma^5 \over 2} (\psi_\uparrow (x^\mu) -\psi_\downarrow^c (x^\mu))\, .
\label{nu2}
\end{equation}
Other fields which we use in order to obtain dynamical equations
are
\begin{mathletters}
\begin{eqnarray}
\nu^c (x^\mu) &=& - {1\over 2}
(\psi_\uparrow (x^\mu) - \psi_\downarrow^c (x^\mu)) + {\gamma^5 \over 2}
(\psi_\downarrow (x^\mu) +\psi_\uparrow^c (x^\mu) )\, ,\\
\gamma^5\nu (x^\mu) &=& - {1\over 2}
(\psi_\uparrow (x^\mu) + \psi_\downarrow^c (x^\mu)) + {\gamma^5 \over 2}
(\psi_\downarrow (x^\mu) -\psi_\uparrow^c (x^\mu) )\, ,\\
\gamma^5\nu^c (x^\mu) &=& + {1\over 2}
(\psi_\downarrow (x^\mu) + \psi_\uparrow^c (x^\mu)) - {\gamma^5 \over 2}
(\psi_\uparrow (x^\mu) -\psi_\downarrow^c (x^\mu) )\, ,\\
\widetilde\nu^c (x^\mu) &=&  +{1\over 2}
(\psi_\uparrow (x^\mu) + \psi_\downarrow^c (x^\mu)) + {\gamma^5 \over 2}
(\psi_\downarrow (x^\mu) -\psi_\uparrow^c (x^\mu) )\, ,\\
\gamma^5\widetilde\nu (x^\mu) &=& + {1\over 2}
(\psi_\uparrow (x^\mu) - \psi_\downarrow^c (x^\mu)) + {\gamma^5 \over 2}
(\psi_\downarrow (x^\mu) + \psi_\uparrow^c (x^\mu) )\, ,\\
\gamma^5\widetilde\nu^c (x^\mu) &=&  + {1\over 2}
(\psi_\downarrow (x^\mu) - \psi_\uparrow^c (x^\mu)) + {\gamma^5 \over 2}
(\psi_\uparrow (x^\mu) +\psi_\downarrow^c (x^\mu) )\, .
\end{eqnarray}
\end{mathletters}
After rather tiresome calculational procedure
one obtains the dynamical equations in this approach
\begin{mathletters}
\begin{eqnarray}
i\gamma^\mu \widetilde \nabla_\mu (\nu -\nu^c) -e\gamma^\mu \gamma^5 C_\mu
(\nu -\nu^c) -m \gamma^5 (\widetilde \nu + \widetilde \nu^c) &=& 0\,
,\\
i\gamma^\mu \widetilde \nabla_\mu (\widetilde \nu + \widetilde \nu^c)
+e\gamma^\mu \gamma^5 C_\mu (\widetilde \nu + \widetilde \nu^c) +m
\gamma^5 (\nu - \nu^c) &=& 0\, ,\\
i\gamma^\mu \gamma^5\widetilde \nabla_\mu (\nu -  \nu^c) -e\gamma^\mu
C_\mu (\nu - \nu^c) +m (\widetilde \nu + \widetilde \nu^c)
&=& 0\, ,\\
i\gamma^\mu \gamma^5\widetilde \nabla_\mu (\widetilde \nu +
\widetilde \nu^c) +e\gamma^\mu C_\mu (\widetilde \nu + \widetilde \nu^c)
-m (\nu - \nu^c) &=& 0\, ,
\end{eqnarray} \end{mathletters}
or
\begin{mathletters}
\begin{eqnarray}
(1+\gamma^5) \left [ i\gamma^\mu \widetilde \nabla_\mu (\nu -\nu^c
+\widetilde \nu +\widetilde \nu^c ) +e\gamma^\mu C_\mu
(\nu -\nu^c -\widetilde \nu -\widetilde \nu^c ) +m
(\nu -\nu^c -\widetilde \nu -\widetilde \nu^c ) \right ] &=& 0\, ,\\
(1+\gamma^5) \left [ i\gamma^\mu \widetilde \nabla_\mu  (\nu -\nu^c
-\widetilde \nu -\widetilde \nu^c ) +e\gamma^\mu C_\mu
(\nu -\nu^c +\widetilde \nu +\widetilde \nu^c ) -m
(\nu -\nu^c +\widetilde \nu +\widetilde \nu^c ) \right ] &=& 0\, ,\\
(1-\gamma^5) \left [ i\gamma^\mu \widetilde \nabla_\mu (\nu -\nu^c
+ \widetilde \nu + \widetilde \nu^c ) -e\gamma^\mu C_\mu
(\nu -\nu^c -\widetilde \nu -\widetilde \nu^c ) -m
(\nu -\nu^c -\widetilde \nu -\widetilde \nu^c ) \right ] &=& 0\, ,\\
(1-\gamma^5) \left [ i\gamma^\mu \widetilde \nabla_\mu (\nu -\nu^c
- \widetilde \nu - \widetilde \nu^c ) -e\gamma^\mu C_\mu
(\nu -\nu^c +\widetilde \nu +\widetilde \nu^c ) +m
(\nu -\nu^c +\widetilde \nu +\widetilde \nu^c ) \right ] &=& 0\, .
\end{eqnarray} \end{mathletters}
Thus, one can see the operators
$$\nu (x^\mu) -\nu^c (x^\mu) \,\, ,\,\, \mbox{and} \,\,\,
(\widetilde \nu (x^\mu) +\widetilde \nu^c (x^\mu) )\, .$$
also satisfy the equations of the type (\ref{two1},\ref{two2}).
From the formulas (\ref{nu1},\ref{nu2}) one can figure out, why do Eqs.
(\ref{110},\ref{120}) and (\ref{1100}-\ref{1400}) have different forms
and how are ${\cal R}^{S,A}$, ${\cal L}^{S,A}$, the self/anti-self charge
conjugate operators, and $\lambda^{S,A}$, $\rho^{S,A}$,
the operators answering for the self/anti-self charge conjugate states,
connected?

\section{The term ${\sigma}\cdot \left [ {\bf A}
\times {\bf A}^\ast  \right ] $:  To Be or Not To Be?}

The possibility of terms as $\sim \bbox{\sigma}\cdot [{\bf A}\times {\bf
A}^\ast]$, ref.~\cite{Evans,Espos}, appears to be related to the matters of
chiral interactions.  As we are now convinced, the Dirac field operator
can be always presented as a superposition of the self- and anti-self
charge conjugate `field operators'. The anti-self charge conjugate part
can give the self charge conjugate part after multiplying by the
$\gamma^5$ matrix and {\it vice versa}.  In the equations (9) we are able
to put further constraints to extract the self-conjugate states. For
instance,
\begin{equation} \gamma^5 \psi_2 =\psi_1 \end{equation} or, the
anti-self charge conjugate states:
\begin{equation} \gamma^5 \psi_1 =
-\psi_2\,.  \end{equation}
Hence, one has\footnote{The anti-self charge
conjugate field function $\psi_2$ can also be used. The equation has then
the form:
\begin{equation} [i\gamma^\mu D_\mu^\ast + m ] \psi_2^a = 0\, .
\end{equation}
}
\begin{equation}
[i\gamma^\mu D_\mu^\ast - m ] \psi_1^s = 0\, ,\label{des}\\
\end{equation}
or\footnote{The self charge conjugate
field function $\psi_1$ also can be used. The equation has
the form:
\begin{equation}
[i\gamma^\mu D_\mu + m ] \psi_1^s = 0\, .
\end{equation}
As readily seen in the cases of alternative choices we have
opposite ``charges" in the terms of the type $\sim \bbox{\sigma}
\cdot [{\bf A}\times {\bf A}^\ast]$ and in the mass terms.}
\begin{equation}
[i\gamma^\mu D_\mu - m ] \psi_2^a = 0\, \label{dea},
\end{equation}
Both equations lead to the terms of interaction such as $\sim
\bbox{\sigma}\cdot [{\bf A} \times {\bf A}^\ast ]$
provided that the 4-vector potential is considered as a complex
function(al). In fact, from (\ref{des}) we have:
\begin{mathletters}
\begin{eqnarray}
i\sigma^\mu \nabla_\mu
\chi_1 -m \phi_1 &=& 0\, \\
i\widetilde\sigma^\mu \nabla_\mu^\ast \phi_1 -m
\chi_1 &=& 0\, .
\end{eqnarray}
\end{mathletters}
And, from (\ref{dea}) we have
\begin{mathletters}
\begin{eqnarray}
i\sigma^\mu \nabla_\mu^\ast \chi_2 -m \phi_2 &=& 0\, \\
i\widetilde\sigma^\mu \nabla_\mu \phi_2 -m \chi_2 &=& 0\,
\end{eqnarray}
\end{mathletters}
The meanings of $\sigma^\mu$ and $\widetilde\sigma^\mu$ are obvious from
the definition of $\gamma$ matrices. From the above set we extract
the terms as $\pm e^2 \sigma^i_{Pauli} \sigma^j_{Pauli} A_i A_j^\ast$,
which lead to the discussed terms~\cite{Evans,Espos}.

Furthermore, one can come to the same conclusions
not applying to the constraints on the creation/annihilation
operators (which we chosen previously for clarity and simplicity).
It is possible to work with self/anti-self charge conjugate fields
${\cal R}^{S,A}$ and ${\cal L}^{S,A}$ and {\it two} Majorana
{\it anzatzen}, see equations (\ref{m1},\ref{m2}).

Thus, in the considered cases it is the $\gamma^5$ transformation
which distinguishes various field configurations (helicity,
self/anti-self charge conjugate properties etc) in the coordinate
representation.

It would be interesting to compare the above arguments for derivation of
the Esposito-Recami-Evans term with those which have been used
in~\cite{Espos}. We would like to note that in the submitted
Esposito-Recami paper the terms of the type $\sim \bbox{\sigma}\cdot
[{\bf A}\times {\bf A}^\ast]$ can be reduced to $(\bbox{\sigma}\cdot
\nabla )  {\cal V}$, where ${\cal V}$ is the scalar potential.

\section{Generalizations to Higher Spin Representations and Conclusions}

As we have learnt the $\gamma^5$ interactions is intimately related to
the question of defining the self/anti-self charge conjugate states.
But, as we discussed in~[9c] (see also~\cite{DVA}) it is impossible to
introduce self/anti-self charge conjugate momentum-space objects in the
$(1,0)\oplus (0,1)$ representation. One can see the problems of
introducing analogues of $\psi_{1,2,3,4}$ in the $(1,0)\oplus (0,1)$
representation, for instance, from these formulas:
\begin{mathletters}\begin{eqnarray}
\psi_\uparrow (x^\mu) &=& \int \frac{d^3 {\bf p}}{(2\pi)^3 2E_p}
\left [ u_\uparrow (p^\mu) a_\uparrow (p^\mu) e^{-i\phi}
+ {\cal C} u_\uparrow^\ast  (p^\mu) b_\downarrow^\dagger
(p^\mu) e^{+i\phi} \right ] \,\, ,\\
\psi_\downarrow (x^\mu) &=& \int \frac{d^3 {\bf p}}{(2\pi)^3 2E_p}
\left [ u_\downarrow (p^\mu) a_\downarrow (p^\mu) e^{-i\phi}
+ {\cal C} u_\downarrow^\ast  (p^\mu) b_\uparrow^\dagger (p^\mu)
e^{+i\phi} \right ]\, \, ,\\
\psi_\rightarrow (x^\mu) &=& \int \frac{d^3 {\bf p}}{(2\pi)^3 2E_p}
\left [ u_\rightarrow (p^\mu) a_\rightarrow (p^\mu) e^{-i\phi}
- {\cal C} u_\rightarrow^\ast  (p^\mu) b_\rightarrow^\dagger (p^\mu)
e^{+i\phi} \right ]\, \, ,
\end{eqnarray}\end{mathletters}
and
\begin{mathletters}\begin{eqnarray}
\psi_\uparrow^c (x^\mu) &=& \int \frac{d^3 {\bf p}}{(2\pi)^3 2E_p}
\left [ -u_\uparrow (p^\mu) b_\downarrow (p^\mu) e^{-i\phi}
+ {\cal C} u_\uparrow^\ast  (p^\mu) a_\uparrow^\dagger
(p^\mu) e^{+i\phi} \right ] \,\, ,\\
\psi_\downarrow (x^\mu) &=& \int \frac{d^3 {\bf p}}{(2\pi)^3 2E_p}
\left [ -u_\downarrow (p^\mu) b_\uparrow (p^\mu) e^{-i\phi}
+ {\cal C} u_\downarrow^\ast  (p^\mu) a_\downarrow^\dagger (p^\mu)
e^{+i\phi} \right ]\, \, ,\\
\psi_\rightarrow (x^\mu) &=& \int \frac{d^3 {\bf p}}{(2\pi)^3 2E_p}
\left [ u_\rightarrow (p^\mu) b_\rightarrow (p^\mu) e^{-i\phi}
+ {\cal C} u_\rightarrow^\ast  (p^\mu) a_\rightarrow^\dagger (p^\mu)
e^{+i\phi} \right ]\, \, ,
\end{eqnarray}\end{mathletters}
The equation $S^c_{[1]} \psi (x^\mu) = e^{i\alpha} \psi (x^\mu)$ with
\begin{equation}
S^c = {\cal C}{\cal K}= e^{i\vartheta^c_{[1]}}\pmatrix{0&\Theta_{[1]}\cr
-\Theta_{[1]}&0\cr}{\cal K}\,\,,\quad \Theta_{[1]} =\pmatrix{0&0&1\cr
0&-1&0\cr
1&0&0\cr}
\end{equation}
defined in ref.~\cite{BWW,DVA}, has no solutions in the field of complex
numbers. We use the following analogues of the formulas of the footnote 2
for the Barut-Muzinich-Williams matrices
\begin{mathletters}
\begin{eqnarray}
&&{\cal C}^{^T}=-{\cal C}\,\,,\quad {\cal C}^\ast={\cal C}=
-{\cal C}^{-1}\\
&&{\cal C} \gamma^{\mu\nu^{\ast}} {\cal C}^{-1} -\gamma^{\mu\nu}\,\,,
\quad {\cal C} \gamma^{5^\ast} {\cal C}^{-1} =-\gamma^5\,\, .
\end{eqnarray}
\end{mathletters}
Furthermore, in the Majorana representation of $\gamma^{\mu\nu}$ matrices
the operator of the charge conjugation ($\vartheta_c =0$) is equal to
\begin{equation}
S^c_{[1]} = \gamma^5_{MR} {\cal K} = i\gamma^5_{WR} \gamma^0_{WR}
{\cal K} =\pmatrix{0&i\cr -i&0\cr} {\cal K}\,\,.
\end{equation}

Thus, if one implies that $\wp_{u,v} =i (\partial/\partial t)/E =\pm 1$
the Weinberg-Ahluwalia equation of ref.~\cite{BWW} transforms as follows:
\begin{equation}
[\gamma^{\mu\nu} \partial_\mu \partial_\nu +\wp_{u,v} m^2 ] \psi (x^\mu) =0
\Rightarrow
[\gamma^{\mu\nu} \partial_\mu \partial_\nu +\wp_{u,v} m^2 ] \psi^c (x^\mu)
=0\, .
\end{equation}
Please notice that the operator $\wp_{u,v}=\pm 1$ defined above
has the following property with respect to the ordinary complex
conjugation $\wp_{u,v}^\ast = -\wp_{u,v}$, cf.~\cite{Sant}.

Finally, in the Majorana representation the analogues of
(\ref{eq1}-\ref{eq4}) have the form:
\begin{mathletters} \begin{eqnarray}
&&\left [ \gamma^{\mu\nu} \nabla_\mu \nabla_\nu + \wp_{u,v} m^2 \right ]
\psi (x^\mu) = 0\,,\\
&&\left [ \gamma^{\mu\nu} \nabla_\mu^\ast \nabla_\nu^\ast + \wp_{u,v} m^2
\right ] \gamma^5 \psi^\ast (x^\mu) = 0\, ,\\
&&\left [ \gamma^{\mu\nu} \nabla_\mu
\nabla_\nu -\wp_{u,v} m^2 \right ] \gamma^5 \psi (x^\mu) = 0\,,\\
&&\left [\gamma^{\mu\nu} \nabla_\mu^\ast \nabla_\nu^\ast  -\wp_{u,v} m^2
\right ] \psi^\ast (x^\mu) = 0\, .
\end{eqnarray}
\end{mathletters}
It is seen
after calculations which are similar to  above that the combinations
$\psi+\psi^\ast$ and $\gamma^5 (\psi - \psi^\ast)$ may have interaction
of the `axial' form.

Concluding, we state that the $\gamma^5$ interaction is indispensable
element of $(1/2,0)\oplus (0,1/2)$ representation space
(and, presumably, of all the representations of the type $(j,0)\oplus
(0,j)$).  We discussed important physical consequences of the presence of
this interaction for the particles of this representation and found
relations to other models.  The proper account of such terms may lead to
deeper understanding of the nature of particle interactions in the modern
gauge theories, of the structure  of the Fock space and reasons for
introduction of the latter as well.

{\it Acknowledgments.}
I am grateful to Profs. D. V. Ahluwalia, A. E. Chubykalo,
S. Esposito, A. F. Pashkov, E. Recami, R. M. Santilli and G. Ziino for
discussions and useful information.  Zacatecas University, M\'exico, is
thanked for awarding the professorship.  This work has been partly
supported by the Mexican Sistema Nacional de Investigadores, the Programa
de Apoyo a la Carrera Docente and by the CONACyT, M\'exico under the
research project 0270P-E.

\end{document}